\documentstyle[11pt,newpasp,twoside,epsf]{article}
\def\edcomment#1{\iffalse\marginpar{\raggedright\sl#1\/}\else\relax\fi}
\reversemarginpar
\def\d{\delta}

\def\Mpc{h^{-1} \, \mbox{Mpc}}

\def\etal{{\it et al. }}
\def\spose#1{\hbox to 0pt{#1\hss}}
\def\simlt{\mathrel{\spose{\lower 3pt\hbox{$\mathchar''218$}}
     \raise 2.0pt\hbox{$\mathchar''13C$}}}
\def\simgt{\mathrel{\spose{\lower 3pt\hbox{$\mathchar''218$}}
     \raise 2.0pt\hbox{$\mathchar''13E$}}}
\def\beq{\begin{equation}}
\def\eeq{\end{equation}}
\def\bce{\begin{center}}
\def\ece{\end{center}}
\def\bea{\begin{eqnarray}}
\def\eea{\end{eqnarray}}
\def\ben{\begin{enumerate}}
\def\een{\end{enumerate}}

\def\nn{\nonumber}

\def\brr{\begin{array}}
\def\err{\end{array}}

\newcommand{\xibar}{\overline{\xi}}

\font\twelveBF=cmmib10 scaled 1000

\newcommand{\x}{\hbox{\twelveBF x}}

\newcommand{\lexp}{\mathop{\bigl\langle}}
\newcommand{\rexp}{\mathop{\bigr\rangle}}
\newcommand{\rexpc}{\mathop{\bigr\rangle_c}{}}

\begin{document}

\title{Large Scale Structure in the weakly non-linear regime} 
 \author{E.Gazta\~naga}


\affil{ INAOE, Astrofisica, Apdo Postal 216 y 51, 7200, Puebla, Mexico \\
IEEC/CSIC, Gran Capit\'an 2-4, 08034 Barcelona, Spain}

\begin{abstract}

Is gravitational growth responsible for the observed large scale structure 
in the universe? 
Do we need non-gaussian
initial conditions or non-gravitational physics to explain the large 
scale features traced by galaxy surveys? I will briefly revise the basic 
ideas of non-linear perturbation theory (PT) as a tool to 
understand structure formation,  in particular through the study of 
higher order statistics,  like the skewness and the 3-point function. 
Contrary to what happens with the
second order statistics (the variance or power-spectrum), this test
of gravitational instability is independent of
the overall amplitude of fluctuations and of 
cosmic evolution, so that it
does not require comparing the clustering at different redshifts.
Predictions from weakly non-linear PT
have been compared with observations to place constraints on
our assumptions about structure formation, the initial conditions
and how galaxies trace the mass. 

\end{abstract}


\section{Introduction}

Where does structure in the Universe come from? 
The current paradigm is that it comes from gravitational growth of
some small initial fluctuations. The self-gravity 
of an initial overdensed region 
increases its density contrast so that eventually the
region collapses. For a flat Universe in the linear regime, 
the local density contrast $\delta\equiv \rho/\bar{\rho}-1$ 
grows as the expansion factor, eg $D=a$,
so that since decoupling linear gravitational growth has the potential 
of amplifying fluctuations by at least a factor of a thousand. 
But Gravity is not
linear and when objects start collapsing the growth could be much larger.
On galactic scales one also has to consider other
forces such as hydrodynamics, heating and cooling by friction,
dissipation, feedback mechanism from stars, such as nova and
supernova explosions, interaction with the CMB and so on.

To test if the above picture of gravitational
growth is correct we need to deal with a classical initial condition problem. 
Because gravitational time scales are very slow, we have no way to measure
the growth of individual large scale structures and we need to
resort to the statistical study of mean quantities.
 One can imagine, for example, measuring the
rms fluctuations (at a given scale) at different cosmic times to see if this 
agrees with the predicted amount of gravitational growth, $D$. Observationally
this corresponds to finding the clustering properties of some
tracer of structure (eg galaxies) at different redshifts. 
If the tracer is not perfect,  we will have some statistical {\it biasing}.
The problem with this
approach is that by the time the rms fluctuations change significantly 
there typically has also been a substantial cosmic evolution of 
the corresponding tracers.
Thus, it is difficult to disentangle the effects of the underlying
cosmological model (which sets the rate $D$ of gravitational growth)
from galaxy evolution.

It is therefore important to have a way of testing the gravitational
growth paradigm at a single cosmic time or redshift.
Higher order correlations and weakly non-linear clustering allows us to
do just this. This is because one can construct ratios of higher order
correlations to powers of the two point amplitude which are independent
of cosmic time or cosmological parameters, but still contain information
of the underlying dynamics.

\section{Gravitational Growth in the weakly non-linear regime}

Gravitational growth increases the density contrast 
of initially small fluctuations so that eventually the
region collapses. The details of this collapse depends on the initial
density profile. As an illustration we will focus in the spherically symmetric 
case. Thus we will study  structure growth in the context of 
matter domination, the fluid limit and the shear-free or spherical collapse 
approximation. These turns out to be very good approximation for the 
one-point cumulants of the density fluctuations.
It is then easy to find (see Peebles 1980, Gazta\~naga \& Lobo 2000) the
 following second order differential equation for the density contrast
$\delta$ in the Einstein-deSitter universe ($\Omega_k=\Omega_\Lambda=0$), eg
$a(t) = (t/t_0)^{2/3}$:
\beq
{d^2 \delta\over{d^2\eta}} + {1\over{2}} ~ {d \delta \over{d\eta}} - {3\over{2}}~\delta = 
{4\over{3}}\,{1\over 1+\delta}\,\left({d\delta\over{d\eta}}\right)^2
~+~ {3\over{2}}~\delta^2 \nn
\eeq
where we have shifted to the rhs all non-linear terms, and used the
$\eta\equiv\ln(a)$ as our time variable. This equation 
reproduces the equation of the {\em spherical
collapse} model (SC).  
As one would expect,
this yields a {\em local} evolution so that the
evolved field at a point is just given by a (non-linear)
transformation of the initial field at the same point,
with independence of the surroundings.
The linear solution factorizes the spatial and temporal part:

\beq
\delta_l(\x,t) = \,D(t-t_0) \, \delta(\x,t_0) = \,D(t) \, \delta_0(\x)
\label{deltal} \nn
\eeq
where  $D$ is the {\em linear growth factor}, which from the
above differential equation:
\beq
D(t) = C_1~e^\eta  ~+~ C_2~e^{-3/2\eta} = C_1~ a(t) ~+~ C_2~ a(t)^{-3/2} \nn
\eeq
with growing $D \sim a$ and the decaying
$D\sim a^{-3/2}$ modes.  Thus,
the initial fluctuations, no matter of what amplitude, grow by the same
factor, $D$, and the statistical properties of the initial field are just
linearly scaled. For example the linear rms fluctuations $\sigma_l$ or its
variance $\sigma^2_l$ gives:

\beq
\sigma^2_l ~\equiv~ \xibar_2  ~\equiv~ \lexp \delta^2(t)\rexp
= \lexp D(t-t_0)^2 \delta_0^2 \rexp = D(t-t_0)^2 ~\sigma_0^2 \nn
\eeq
where $\delta_0=\delta(t_0)$ and $\sigma_0$ refer to some initial reference time $t_0$:
$\sigma_0^2 \equiv <\delta_0^2>$. 
 
We are interested in the perturbative regime 
($\delta \rightarrow 0$), which is the relevant one for the
description of structure formation on large scales.
The non-linear solution for $\delta$ can then be expressed directly in terms of the linear
one, $\delta_l$:

\beq
\delta  =  f(\delta_l) =  
\,\sum_{n=1}^{\infty} ~{\nu_n \over n!}\, ~{[\delta_l]^n} \nn
\label{loclag} 
\eeq
Thus all non-linear information is encoded 
in the $\nu_n$ coefficients.
We can now introduce this expansion in our non-linear differential
equation, with $\delta_l$ given by the linear growth factor $D=a=e^\eta$,
and compare order by order to find: 
\beq
\nu_2 = {34\over 21} ~~~;~~~
\nu_3 = {682\over 189} ~~~;~~~
\nu_4 = {446440\over 43659}  ~~~;~~~
\nu_5 = {8546480\over 243243} ~~~~~~ \dots \nn
\label{nusc}
\eeq
These results are  derived for the 
Einstein-de Sitter, but are also 
a good approximation for other cosmologies
(eg Bouchet et al. 1992,  Bernardeau 1994a, 
Fosalba \& Gazta\~naga 1998b, Kamionkowski \& Buchalter 1999).
For non-standard cosmologies or a different equation
of state see Gazta\~naga \& Lobo (2000).

One can now find the N-order cumulants
$\xibar_N \equiv \lexp \delta^N \rexpc$, where $N=2$ corresponds
to the variance. 
Here the expectation values $\lexp ... \rexp$ correspond to
an average over realizations of the initial field. On comparing
with observations we assume the  {\it fair sample hypothesis}
(\S 30 Peebles 1980), by which we can commute spatial integrals
with expectation values. Thus, in practice  $\lexp ... \rexp$
is the average  over positions in the survey area. 
It is useful to introduce the N-order {\it hierarchical 
coefficients}: $S_{N}\,=\xibar_N/\xibar_2^{N-1}$, eg
{\it skewness} for $S_3$ and {\it kurtosis} for $S_4$.
These can easily be estimated  from the series expansion above
by just taking expectation values of different powers of $\delta$
(eg see Fosalba \& Gazta\~naga 1998a). 
For leading order Gaussian initial conditions we have:
\beq
S_{3} = 3~ \nu_2  ~~~;~~~
S_{4} = 4~ \nu_3 + 12~ \nu_2^2   ~~~;~~~
S_{5} = 5~ \nu_4 + 60~ \nu_3 \nu_2+ 60~ \nu_2^3   \nn
\label{s3nu2}
\eeq
These results have also been extended to the non-Gaussian case, see  Fry \& Scherrer (1994)
Chodorowski \& Bouchet (1996), 
Gazta\~naga \& Mahonen (1996),
Gazta\~naga \& Fosalba (1998).
If we take for $\nu_2$ the non-linear solution above, eg $\nu_2=34/21$,
the skewness yields  $S_3= 3 \nu_2= 34/7$, which reproduces the exact 
perturbation theory (PT) result by Peebles (1980).
Thus the above (SC) model gives the exact leading order result
for the skewness. This is also true for higher orders (see Bernardeau
1992 and Fosalba \& Gazta\~naga 1998a). These expressions have to
be corrected for smoothing effects (see Juszkiewicz \etal 1993, 
Bernardeau 1994a, 1994b,  and Fosalba \& Gazta\~naga 1998a) and
possibly from redshift space distortions (eg Hivon et al 1995,
Scoccimarro, Couchman and Frieman 1999).
Next to leading order terms have been estimated by 
Scoccimarro \& Frieman (1996),
Fosalba \& Gazta\~naga (1998a,b). 

The 1-point cumulants measured in galaxy catalogues have been 
compared with these PT predictions (eg Bouchet \etal 1993, 
Gazta\~naga 1992, Gazta\~naga \& Frieman 1994,
Baugh, Gazta\~naga \& Efstathiou 1995, Baugh \& Gazta\~naga 1996, 
Colombi etal 1997, Hui \& Gazta\~naga 1999).
The left panel of Figure 1 shows a comparison of $S_N$ measure in the APM 
Galaxy Survey (Maddox et al. 1990), with the predictions above (see 
Gazta\~naga 1994, 1995 for more details). The agreement between predictions
(lines) and measurements (points) on scales $R > 10 \Mpc$ (where
$\xibar_2<1$), is quite good indicating
that the APM galaxies follow the  non-linear gravitational 
growth picture. For errors on statistics  
see Szapudi, Colombi and Bernardeau (1999) and references therein.

\begin{figure}
\plottwo{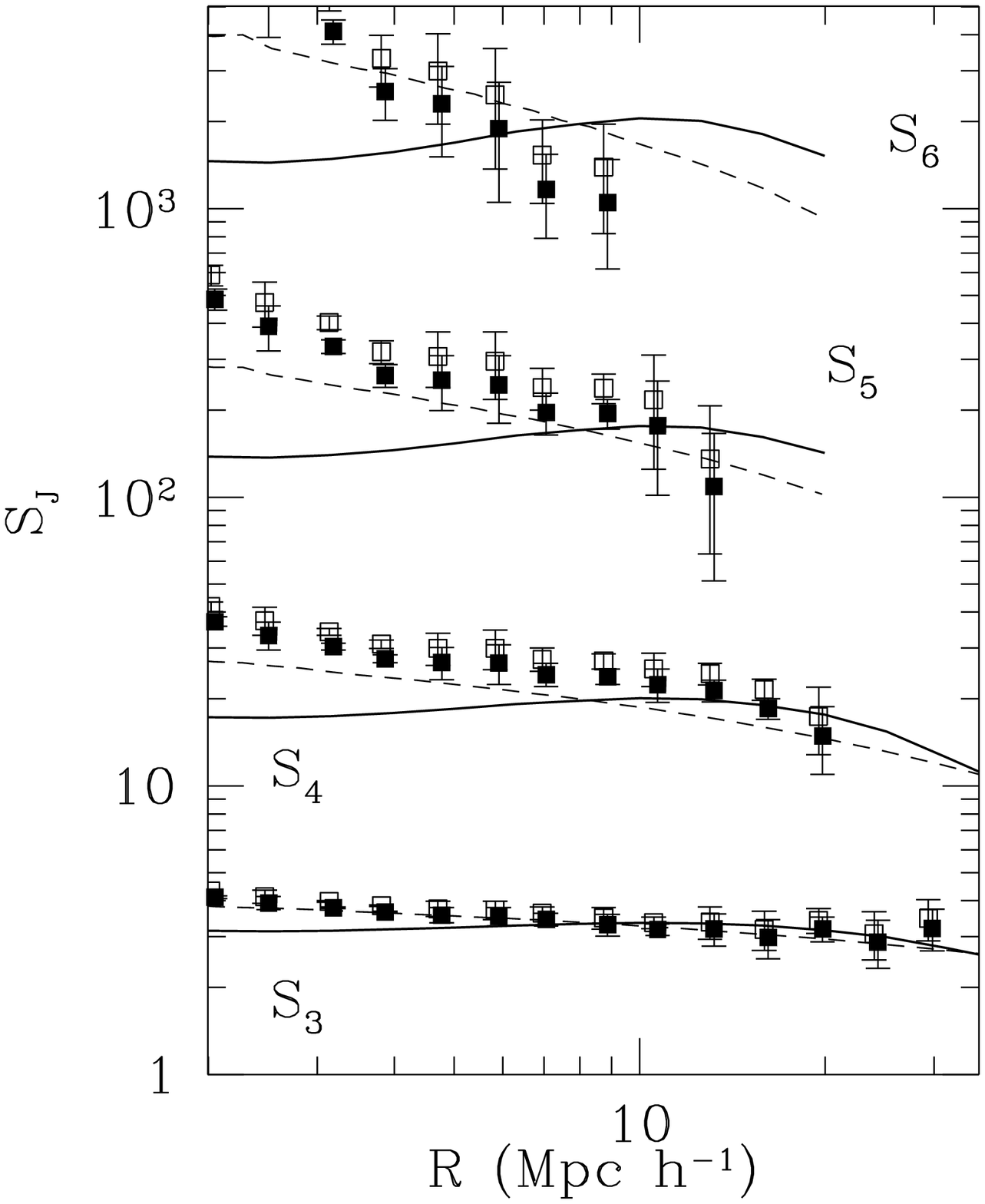}{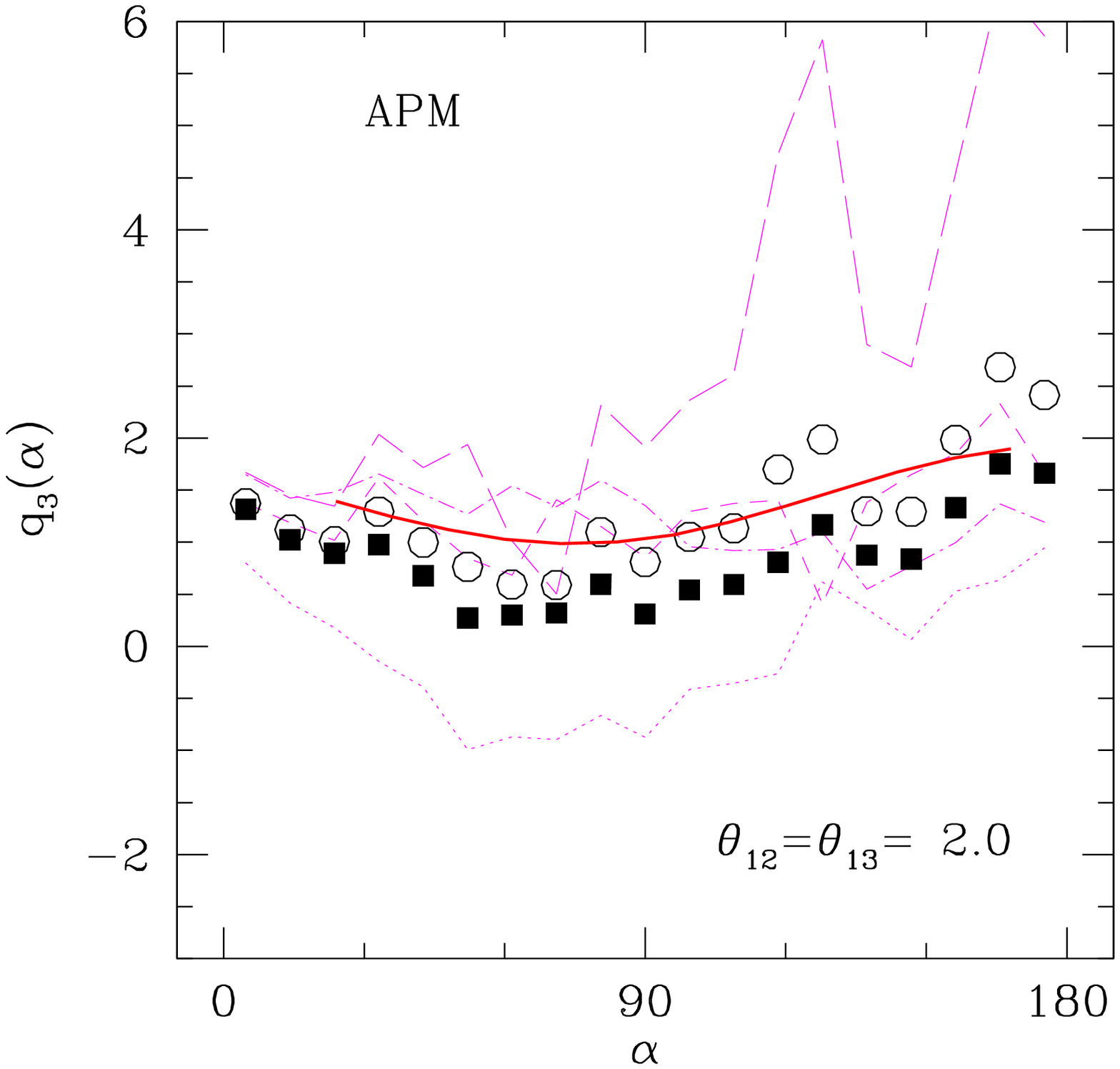}
\caption{Figures in the left panel show  estimates of 
$S_N$, for $N=3-6$, from the APM galaxy catalogue.  
Lines correspond to the PT predictions.
Figures in the right panel shows the APM
 projected 3-point amplitude $q_3(\alpha)$,
for triplets  of objects $1, 2, 3$ on the sky 
at fixed  $\theta_{12}=\theta_{13}=2$ deg and 
as  a function of $\alpha$ the interior angle between these 
two triangle sides. The thick line shows the PT predictions.}
\label{q3nbodyr20}
\end{figure}

\subsection{Biasing: tracing the mass}

The expressions above apply to unbiased tracers of the 
density field; since galaxies of different morphologies are known to 
have different clustering properties, 
at least some galaxy species are biased. As an example, suppose
the probability of forming a luminous galaxy depends  
only on the underlying mean density field in its immediate vicinity. 
The relation between the density field as traced by galaxies
$\d_{gal}(\x)$ and the mass density field $\d(\x)$, can then
be written as:
\beq
\d_{gal}(\x) = f(\d(\x)) = \sum_n ~{b_n \over {n!}} ~~\d^n(\x),  \nn
\eeq
where $b_n$ are the bias parameters. Thus, note how biasing and gravity
could produce comparable non-linear effects. 
To leading order in $\xibar_2$, this local bias  scheme implies $\xibar_2^{gal} =b_1^2 \xibar_2$ and
(see Fry \& Gazta\~naga 1993):  
\beq
S_3^{gal} = {S_3 \over{b_1}} + 3~ {b_2 \over{b^2_1}} ~~~;~~~
S_4^{gal} = {S_4 \over{b_1^2}} + 12~ {b_2 S_3\over b_1^3}  + 4~ {b_3\over b_1^4} + 
12~ {b_2^2\over b_1^4} ~~~;~~~ \dots \nn
\eeq
Gazta\~{n}aga \& Frieman (1994) have used the comparison of $S_3$ and
$S_4$ in PT with the corresponding measured APM values (as shown in Figure 1)
to infer that $b_1 \simeq 1$, $b_2 \simeq 0$ and $b_3 \simeq 0$,
 but the results are 
degenerate due to the relative scale-independence of $S_N$ and the
increasing number of biasing parameters. One could break this
degeneracy by using the configuration-dependence of the projected 3-point function,
$q_3(\alpha)$, as proposed by  Frieman \& Gazta\~naga (1994), Fry (1994),
Matarrese, Verde \& Heavens (1997)
Scoccimarro et al (1998). As shown in Frieman \& Gazta\~naga (1999), the
configuration-dependence of $q_3({\alpha})$ on large scales in the APM 
catalog  is quite close to that expected in perturbation
theory , suggesting again that $b_1$ is of order unity (and $b_2 \simeq 0$)
for these galaxies. This is illustrated in the right panel of Figure 1. 
The solid curves show the predictions of weakly non-linear
gravitational growth. The APM galaxy measurements are shown as  
symbols ; other curves show results for each of the zones.
The agreement indicates that 
large-scale structure is driven by non-linear gravitational 
instability and that APM galaxies are relatively 
unbiased tracers of the mass on these large scales.

\section{Conclusions}

The values of $S_N = \xibar_N/\xibar_2^{N-1}$ 
can be measured as traced by the large scale galaxy distribution
(eg Bouchet \etal 1993, Gazta\~naga 1992, 1994, 
Szapudi el at 1995, Hui \&  Gazta\~naga 1999 and references therein), 
and also the weak-lensing (Bernardeau \etal 1997, 
Gazta\~naga \& Bernardeau 1998) or the Ly-alpha QSO absorptions
(Gazta\~naga \& Croft 1999). These measurements of the skewness $S_3$,
kurtosis $S_4$, and so on, can be 
compared with the predictions
from weakly non-linear perturbation theory (see Figure 1)
to place constraints on
our assumptions about gravitational growth, initial conditions
or biasing at a given redshift (see Mo, Jing \& White 1997). 
Contrary to what happens with the
second order statistics (eg the variance), this test
of gravitational instability is quite independent of
the overall amplitude of fluctuations and
other assumptions of our model for cosmological evolution, and
does not require comparing the clustering at different redshifts.
As shown in Gazta\~naga \& Lobo (2000),
one can also use the $S_N$ measurements to constraint non-standard
cosmologies.

Frieman \& Gazta\~naga (1999) have presented new
results for the angular 3-point galaxy correlation 
function in the APM Galaxy Survey and its comparison with theoretical 
expectations (see also Fry 1984, Scoccimarro et al. 1998,
Buchalter, Jaffe \& Kamionkowski 2000). 
For the first time, these measurements 
extend to sufficiently large scales to probe the weakly non-linear 
regime (see previous work by Groth \& Peebles 1977, Fry \& Peebles 1978,
Fry \& Seldner 1982).
On large scales, the results are in good agreement with the 
predictions of non-linear perturbation theory,  
for a model with initially Gaussian fluctuations 
(see Figure 1). This reinforce the conclusion that 
large-scale structure is driven by non-linear gravitational 
instability and that APM galaxies are relatively 
unbiased tracers of the mass on large scales; they also 
provide stringent constraints upon 
models with non-Gaussian initial conditions 
(eg see Gazta\~naga \& Mahonen 1996; Peebles 1999a,b; White 1999;
Scoccimarro 2000).





\end{document}